\documentclass[aps,tightenlines,showpacs,preprintnumbers]{revtex4}
\usepackage{graphicx}
\usepackage{float}
\usepackage[T1]{fontenc}
\usepackage[latin1]{inputenc}
\usepackage{amssymb}

\newcommand{\bea}{\begin{eqnarray}}
\newcommand{\eea}{\end{eqnarray}}

\makeatother
\preprint{TUW-08-18}
\begin{document}

\title{Next-to-leading order static gluon self-energy for anisotropic plasmas}

\author{M.E. Carrington}

\affiliation{Brandon University,
Brandon, Manitoba, R7A 6A9 Canada\\
and Winnipeg Institute for Theoretical Physics,
Winnipeg, Manitoba, Canada}

\author{A. Rebhan}

\affiliation{Institut f\"{u}r Theoretische Physik, Technische Universit\"{a}t Wien, Wiedner Hauptstra\ss{}e 8-10, A-1040 Vienna, Austria}

\begin{abstract}
In this paper the structure of the next-to-leading (NLO) static gluon self energy for an anisotropic plasma is investigated in the limit of a small momentum space anisotropy. Using the Ward identities for the static hard-loop (HL) gluon polarization tensor and the (nontrivial) static HL vertices, we derive a comparatively compact form for the complete NLO correction to the structure function containing the space-like pole associated with magnetic instabilities. On the basis of a calculation without HL vertices, it has been conjectured that the imaginary part of this structure function is nonzero, rendering the space-like poles integrable. We show that there are both positive and negative contributions when HL vertices are included, highlighting the necessity of a complete numerical evaluation, for which the present work provides the basis.
\end{abstract}


\maketitle

\section{Introduction}

The difficulties in explaining the strong collective phenomena observed at the Relativistic Heavy Ion collider (RHIC) \cite{Tannenbaum:2006ch} by perturbative QCD at finite temperature have led to extensive studies of the consequences of the inevitable presence of non-Abelian plasma instabilities \cite{Mrowczynski:1988dz,Pokrovsky:1988bm,Mrowczynski:1993qm} in a plasma with momentum-space anisotropy. To leading order in the coupling and for small gauge field amplitudes, the dynamics of plasma instabilities are determined by the generalization of the hard-thermal-loop (HTL) \cite{Weldon:1982aq,Frenkel:1990br,Braaten:1990mz} gauge boson self-energy to anisotropic situations \cite{Mrowczynski:2000ed,Romatschke:2003ms,Romatschke:2004jh,Arnold:2003rq}.  

In equilibrium, it is well known that non-bilinear terms in the HTL effective action \cite{Taylor:1990ia,Braaten:1992gm} are important at next-to-leading order (NLO). In order to obtain complete NLO corrections to HTL dispersion laws such as plasmon damping constants, one must resum both propagators and vertices \cite{Braaten:1990it,Braaten:1992gd,Kobes:1992ys,Schulz:1994gf,Carrington:2006gb,Carrington:2008dw}. In anisotropic systems, the non-bilinear terms in the corresponding hard-loop (HL) effective action \cite{Mrowczynski:2004kv} are important for the dynamics of non-Abelian plasma instabilities at large gauge field amplitudes, which has been studied using real-time lattice simulations \cite{Rebhan:2004ur,Arnold:2005vb,Rebhan:2005re,Bodeker:2007fw,Arnold:2007cg,Rebhan:2008uj}. Up to now, analytic calculations at NLO in anisotropic systems have been considered prohibitively difficult. In addition to the obvious technical difficulties associated in dealing with expressions that contain huge numbers of terms, these calculations contain new conceptual problems arising from the fact that the anisotropic gluon propagator contains non-integrable space-like poles. 

In perturbative estimates of jet quenching and momentum broadening in the anisotropic quark-gluon plasma, such non-integrable singularities also appear \cite{Romatschke:2006bb,Baier:2008js}. It has been suggested in \cite{Romatschke:2006bb} that the static gluon self energy might develop a non-zero imaginary part at NLO  that regulates these singularities. In thermal equilibrium, the imaginary part of the static gluon self energy, which is an odd function of the frequency, vanishes due to the KMS condition \cite{LeB:TFT}. In the anisotropic case, it seems possible that there is a finite, discontinuous  contribution. 
Ref. \cite{Romatschke:2006bb} provided support for this conjecture in the form of a partial calculation of the anisotropic NLO static gluon self-energy. This calculation included however only a tadpole diagram with a bare four-vertex. 

\begin{figure}[h]
\begin{center}
\includegraphics[width=10cm]{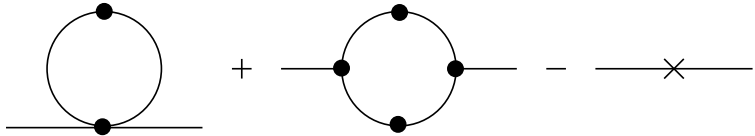}
\end{center}
\caption{The diagrams that contribute to static gluon self energy. All lines correspond to gluon propagators. The dots indicate hard loop propagators and vertices. The cross denotes the HL counterterm.} 
\label{staticSE}
\end{figure}

In this paper, we provide the basis for a complete analytic calculation of the anisotropic  NLO static gluon self-energy. In the HTL case, only static loop momenta have to be considered, no fermion loops contribute, and all HTL vertex corrections vanish \cite{Rebhan:1993az}. However, the anisotropic HL vertex corrections do not vanish in the static limit \cite{Mrowczynski:2004kv}. 
The relevant diagrams are shown in Fig. \ref{staticSE} (since the ghost self energy vanishes at leading order, the ghosts do not need to be resummed).  The solid dots indicate leading order propagators and vertices which are obtained from the hard loop (HL) effective action \cite{Mrowczynski:2004kv}. The third diagram is the HL counterterm which must be subtracted to avoid double counting.

In this paper, we give an analytic result for the integrand corresponding to the diagrams in Fig. \ref{staticSE}.  After extensive algebraic manipulations, the final expression has a relatively simple form, because of cancellations that occur between different contributions from the tadpole and bubble contributions. We divide the result into `tadpole-like' contributions (which contain only one HL propagator) and `bubble-like' contributions (which contain two HL propagators).
We calculate the contribution from `tadpole-like' contributions with a bare vertex. We compare our result with the (corrected) result of \cite{Romatschke:2006bb} for the tadpole diagram (without taking into account cancellations with the bubble diagram) with a bare vertex. 

\par

\section{Notation and hard anisotropic loops}

In this section we define our notation for the anisotropic HL quantities
and give the explicit form of the static HL propagator in Feynman gauge.

At zero temperature, field theory can be formulated covariantly. At finite temperature, covariance is broken by the vector $u_\mu=(1,0,0,0)$ which specifies the rest frame of the thermal system. For anisotropic systems we need (in the simplest case) one additional vector to specify the direction of the anisotropy. We consider the case in which there is one preferred spatial direction along which the system is anisotropic (in planes transverse to this vector the system is isotropic). In the context of heavy ion collisions, we can take this direction to be the beam axis ($\hat z$) along which the initial expansion occurs. 

\subsection{Distribution Functions}

We define the isotropic particle distribution:
\bea
\label{f-iso}
f_{\rm iso}(p)=2 N_f[n(p)+\tilde n(p)]+4 N_c n^g(p)
\eea
In equilibrium, we have
\bea\label{neq}
n_{\rm eq}(p)=\frac{1}{e^{(p-\mu)/T}+1}\,;~~\tilde n_{\rm eq}(p)=\frac{1}{e^{(p+\mu)/T}+1}\,;~~n^g_{\rm eq}(p)=\frac{1}{e^{p/T}-1}
\eea
We define the Debye mass from the equilibrium distribution$^1$
\footnotetext[1]{In Ref. \cite{Romatschke:2003ms,Romatschke:2004jh} Eq. (\ref{debye}) differs by a factor of 2 and  Eq. (\ref{f-iso}) differs by a factor 1/2. The definition of the Debye mass is the same.}
\bea
\label{debye}
m_D^2=g^2 \int \frac{d^3 p}{(2\pi)^3} \,\frac{f_{eq}(p)}{p} = -\frac{g^2}{2} \int \frac{d^3 p}{(2\pi)^3} \, \frac{d f(p)}{d p} = \frac{1}{3}N_c g^2 T^2+\frac{1}{6}N_f g^2 \left(T^2+\frac{3}{\pi^2}\;\mu^2\right)\,.
\eea

Following \cite{Romatschke:2003ms,Romatschke:2004jh}, we can construct an anisotropic distribution from any isotropic distribution of the form $f_{\rm iso}(p^2)$ by writing
\bea
f(\vec p) = f_{\rm iso}(p^2+\xi(\vec p\cdot \hat z)^2)
\eea
where $\xi > -1$ is the anisotropy parameter. A value $\xi>0$ corresponds to a contraction of the distribution and $0>\xi>-1$ corresponds to a stretching of the distribution. In this paper we restrict ourselves to weakly anisotropic systems for which $|\xi|\ll 1$.

\subsection{Projection Operators}

Following \cite{Kobes:1991dc} we construct the vector 
\bea
\label{n-defn}
n^i(k)=(\delta^{ij}-k^ik^j/k^2)\delta^{j3} = \delta^{i3}-k^i k^{(3)}/k^2\,;~~k^{(3)}:=\vec k\cdot \hat z=k^j\delta^{j3}
\eea
which satisfies $n(k)^i \,k^i = 0$.
Throughout this paper we will frequently use the indices $\{k,q,r\}$ to denote momentum arguments.  For example: we will write $n^i_k:= n^i(k)$. We also use Latin letters $\{i,j,l,t,\cdots\}$ to denote spatial indices, with the exception that the indices $k$, $q$, $r$ are reserved and used exclusively to denote momenta. Using the vector in (\ref{n-defn}), we construct the projection operators:
\bea
P^1_{ij}= \delta _{ij}-\frac{k_i k_j}{k^2}\,;~~
P^2_{ij}=\frac{k_i k_j}{k^2}\,;~~
P^3_{ij}= -\frac{k_i k_j}{k^2}-\frac{n^k_i n^k_j}{n_k^2}+\delta _{ij}\,;~~
P^4_{ij}=k_j n^k_i+k_i n^k_j\,;~~
P^5_{ij}=P^1_{ij}-P^3_{ij}
\eea
which satisfy:
\bea
&& P^1_{il}P^1_{lj}=P^1_{ij}\,;~~P^1_{il}P^2_{lj}=0\,;~~P^1_{il}P^3_{lj}=P^3_{ij}\,;~~P^1_{il}P^4_{lj}+P^4_{il}P^1_{lj}=P^4_{ij} \\
&&P^2_{il}P^2_{lj}=P^2_{ij}\,;~~P^2_{il}P^3_{lj}=0\,;~~P^2_{il}P^4_{lj}+P^4_{il}P^2_{lj}=P^4_{ij} \nonumber\\
&&P^3_{il}P^3_{lj}=P^3_{ij}\,;~~P^3_{il}P^4_{lj}=0 \nonumber\\
&&P^2_{il}P^5_{lj}=0\,;~~P^3_{il}P^5_{lj}=0\,;~~P^4_{il}P^5_{lj}=P^4_{ij}\,;~~P^5_{il}P^5_{lj} = P^5_{ij}\nonumber\\
&&k_iP^1_{ij}=k_iP^3_{ij}=k_ik_jP^4_{ij}=k_iP^5_{ij}=0 \nonumber\\[4mm]
&&n^k_iP^1_{ij}=n^k_j\,;~~n^k_iP^2_{ij}=n^k_iP^3_{ij}=0\,;~~n^k_i n^k_j P^4_{ij}=0 \,;~~n^k_i P^5_{ij}=n^k_j\nonumber\\
&& {\rm Tr}\,P^1 = 2\,;~~ {\rm Tr}\,P^2 ={\rm Tr}\,P^3 ={\rm Tr}\,P^5= 1\,;~~{\rm Tr}\,P^4=0\nonumber
\eea

\subsection{Lowest Order Self Energy}

We define the polarization tensor using the relation:
\bea
\label{PIdef}
D^{-1}_{\mu\nu}(K) =(D^{0}_{\mu\nu})^{-1}-\Pi_{\mu\nu} = -(g_{\mu\nu}K^2+\Pi_{\mu\nu})
\eea
The HL gluon self energy is gauge invariant and satisfies the usual Ward identity: $K^\mu \Pi_{\mu\nu}=0$. Consequently, we only need to calculate the spatial components. At finite temperature, there are two independent components which are called the transverse and longitudinal parts. For anisotropic systems the self energy can be decomposed into four independent structure functions:
\bea
\Pi^k_{ij} = P^1_{ij}\alpha_k + P^2_{ij}\,\bar\beta_k + P^5_{ij}\,\gamma_k + P^4_{ij}\,\bar\delta_k
\eea
The structure functions are calculated from:
\bea
\label{pi-invert}
&& x = \mathcal P(x)_{ij}\Pi_{ij}\,;~~x \in \{\alpha,\bar\beta,\gamma,\bar\delta\} \\[4mm]
&& \mathcal P(\alpha)_{ij} = P^3_{ij}\,;~~\mathcal P(\bar\beta)_{ij} = P^2_{ij}\,;~~\mathcal P(\gamma)_{ij} = P^1_{ij}-2P^3_{ij}\,;~~\mathcal P(\bar\delta)=P^4_{ij}/(2k^2 n_k^2)\nonumber
\eea
In the limit of small anisotropy parameter $\xi$,
the lowest order results are \cite{Romatschke:2003ms,Romatschke:2004jh}:
\bea
\label{PIres}
&&\alpha_k =\frac{k_0^2 m_D^2}{k^2}+\xi  \left(\frac{\left(5-7n_k^2\right) k_0^2 m_D^2}{3 k^2}-\frac{1}{3}(1-n_k^2)
   m_D^2\right)
   + i \left(\frac{\left(2-3n_k^2\right) \pi  \xi  k_0 m_D^2}{8 k}+\frac{\pi  k_0 m_D^2}{4 k}\right) \nonumber\\
&& \bar\beta_k =\frac{\left(3n_k^2-1\right) \xi  k_0^2 m_D^2}{3 k^2}-\frac{k_0^2 m_D^2}{k^2}\nonumber\\
&& \gamma_k =\xi  \left(\frac{-4 n_k^2 k_0^2 m_D^2}{3 k^2}+\frac{1}{3} m_D^2 n_k^2\right) - \frac{i n_k^2 \pi  \xi  k_0 m_D^2}{4 k} \nonumber\\
&&\bar\delta_k =\frac{7 k^{(3)} \xi  k_0^2 m_D^2}{3 k^4}+\frac{i k^{(3)} \pi  \xi  k_0 m_D^2}{4 k^3}
   \eea
For convenience we define:
\bea
k_0 \delta_k := \bar\delta_k\,;~~k_0^2 \beta_k := \bar\beta_k
\eea

\subsection{Propagator}

We calculate the static propagator in the covariant gauge (with Feynman gauge parameter) by inverting (\ref{PIdef}).
This inversion is complicated by the fact that the anisotropic propagator depends on the two fixed vectors (1,0,0,0) and (0,0,0,1), as well as the 4-momentum $K_\mu$. The method is described in \cite{Kobes:1991dc}. The result is:
\bea
\label{static-prop}
&& D_{ij}(k) = -P^2_{ij}\frac{1}{k^2}-G_A(k)P^3_{ij}+n_k^2 F_D(k)P^5_{ij}\\
&&  D_{0i}(k) =F_B (n_k)_i\nonumber\\
&& D_{00} = F_C(k)\nonumber\\[4mm]
&& F_B(k) = \delta_k G_B(k)   \nonumber \\
&& k^2 F_C(k) = (k^2+\alpha_k +\gamma_k)G_B(k) \nonumber\\
&& n_k^2 F_D(k) = \beta_k G_B(k)  \nonumber\\[4mm]
&& G_A^{-1}(k) = k^2+\alpha_k \nonumber \\
&& G_B^{-1}(k) = (1-\beta_k)(k^2+\alpha_k+\gamma_k)+k^2 n_k^2 \delta_k^2 \nonumber
\eea

In the small-$\xi$ limit, the result (11) implies that $G_B$ has a pole at negative $k^2=-m_D^2 + {\cal O}(\xi),$ which corresponds to the usual electric (Debye) screening. For non-vanishing $\xi$ space-like poles also appear. Writing $\hat k\cdot \hat z = \cos \theta$, and $n_k^2 = 1-(\hat k\cdot \hat z )^2 = \sin^2\theta,$ we find that these space-like poles appear at:
\bea
&&G_B(k)\,:~~~~k^2/m_D^2=\frac{1}{3}\,\xi\, (1-2n_k^2) = \frac{1}{3}\,\xi\, \cos 2\theta \\
&&G_A(k)\,:~~~~k^2/m_D^2=\frac{1}{3}\,\xi\, (1-n_k^2)=\frac{1}{3}\,\xi\, \cos^2\theta \nonumber
\eea
When $\xi$ is positive, corresponding to an oblate particle distribution, $G_B(k)$ has a space-like pole for $\pi > \theta > 3\pi/4$ and $\pi/4 > \theta > 0$. For $\xi$ negative, the space-like pole occurs for $ 3\pi/4 > \theta > \pi/4$. $G_A(k)$ has a space-like pole for any positive value of $\xi$ unless $\hat k \cdot \hat z = 0$. These poles of $G_A$ correspond to the magnetic Weibel instability \cite{Weibel:1959}.

\subsection{Vertices}

In this section we give our notation for the HL vertices \cite{Mrowczynski:2004kv}. We define:
\bea
&&\int_p:= \frac{d^3p}{(2\pi)^3} \Big|_{p_0=p}\;;~~ \hat P^\mu := (1,\hat p^i)\;;~~
 \hat I_\beta := \frac{g^2}{2}\int_p \frac{\partial f}{\partial P^\beta}
\eea

The 2-point function is:
\bea
&& \Pi_{ab}^{\mu\nu}:=\delta_{ab}\Pi^{\mu\nu} \;;~~\Pi^{\mu\nu} := \hat I_\beta \hat P^\mu \hat\Pi^{\nu\beta}\;;~~ \hat\Pi^{\nu\beta} := \left(g^{\nu\beta}-\frac{\hat P^\nu Q^\beta}{P\cdot Q}\right) 
\eea
The 3-point function is$^2$
\footnotetext[2]{In Ref. \cite{Mrowczynski:2004kv} there is a missing factor (-1) in Eq. (35)}
\bea
&&\Gamma^{\mu\nu\lambda}_{abc}:=i g f_{abc}\Gamma^{\mu\nu\lambda} \\
&& \Gamma^{\mu\nu\lambda} := \hat I_\beta P^{\mu}P^{\nu}P^{\lambda}\hat\Gamma^\beta \nonumber\\
&& \hat\Gamma^\beta := \frac{K^\beta}{\hat P\cdot K\;\hat P\cdot Q} - \frac{R^\beta}{\hat P\cdot Q\;\hat P\cdot R} \nonumber
\eea
The 4-point function is$^3$ \footnotetext[3]{In Ref. \cite{Mrowczynski:2004kv} there is a missing factor (-1/2) in Eq. (39) and a missing $i$ in Eq. (38)}
\bea
&&M^{\mu\nu\lambda\sigma}_{abcd} := -2 g^2 \big(X_{abcd}\,M^{\mu\nu\lambda\sigma}~+~2~{\rm cyclic~permutations}\big)\;;~~X_{abcd} = {\rm tr}\left(T_d[T_c,[T_b,T_a]]\right) \\
&& M^{\mu\nu\lambda\sigma} := \hat I_\beta \hat P^\mu \hat P^\nu \hat P^\lambda \hat P^\sigma \hat M^\beta \nonumber \\
&& \hat M^\beta := -\frac{K^\beta}{\hat P \cdot K \; \hat P \cdot Q\; \hat P\cdot(Q+L)}
-\frac{K^\beta + Q^\beta}{\hat P \cdot Q \; \hat P \cdot L\; \hat P\cdot(L+S)}
-\frac{K^\beta+Q^\beta+L^\beta}{\hat P \cdot L \; \hat P \cdot S\; \hat P\cdot(K+S)}\nonumber
\eea
Momenta are taken to be incoming. When the momentum arguments are not written explicitly, they are taken to be in the order $(K,Q,R=-K-Q)$ for the 3-point function and $(K,Q,L,S=-K-Q-L)$ for the 4-point function.
The HL vertices satisfy the Ward identities:
\bea
\label{wi-full}
&& K_\mu \Gamma^{\mu\nu\lambda}(K,Q,R) = \Pi^{\nu\lambda}(Q) - \Pi^{\nu\lambda}(R) \\
&& K_\mu M^{\mu\nu\lambda\sigma}(K,Q,L,S) = \Gamma^{\nu\lambda\sigma}(Q,L,-L-Q) - \Gamma^{\nu\lambda\sigma}(-L-S,L,S) \nonumber
\eea

The tadpole form of the 4-point vertex has a simpler form:
\bea
&& M^{\mu\nu\lambda\sigma}_{abcc}(Q,-Q,K,-K):=2 g^2\,C_A \delta_{ab} \;M^{\mu\nu\lambda\sigma}(Q,-Q,K,-K) \\
&&M^{\mu\nu\lambda\sigma}(Q,-Q,K,-K) = -2 \hat I_\beta \hat P^\mu \hat P^\nu \hat P^\lambda \hat P^\sigma \left(\frac{K^\beta \hat P \cdot Q- Q^\beta \hat P \cdot K }{P \cdot K\;P \cdot Q\;((P \cdot K)^2-(P \cdot Q)^2)}\right) \nonumber
\eea
This vertex satisfies the Ward identities:
\bea
\label{wi-tad}
K_\lambda M^{\mu\nu\lambda\sigma}(Q,-Q,K,-K) = -2 \Gamma^{\mu\nu\sigma}(K,Q,-K-Q) \\
K_\lambda K_\sigma M^{\mu\nu\lambda\sigma}(Q,-Q,K,-K) = 2 \Pi^{\mu\nu}(-K-Q) - 2 \Pi^{\mu\nu}(Q) \nonumber
\eea

We also define the bare vertices:
\bea
&& (\Gamma_0)_{abc}^{\mu\nu\lambda} = igf_{abc}\Gamma_0^{\mu\nu\lambda} \\
&& \Gamma_0^{\mu\nu\lambda} = -g^{\mu\nu}(K^\lambda-Q^\lambda)-g^{\lambda\nu}(Q^\mu-R^\mu)-g^{\lambda\mu}(R^\nu-K^\nu) \nonumber\\
&& (M_0)_{abcc}^{\mu\nu\lambda\sigma}(Q,-Q,K,-K) = 2 g^2\,C_A \delta_{ab} \;M^{\mu\nu\lambda\sigma}(Q,-Q,K,-K) \nonumber\\
&& M_0^{\mu\nu\lambda\sigma} = -g^{\lambda\nu}g^{\mu\sigma}-g^{\lambda\mu}g^{\nu\sigma}+2g^{\lambda\sigma}g^{\mu\nu}\nonumber
\eea

Using these definitions we write the integrand corresponding to the first two diagrams in Fig. \ref{staticSE} as:
\bea
\label{integrand}
\Pi^{\mu\nu}(Q) = \frac{1}{2} \,g^2\, C_A \,\delta_{ab} \,T\, \int \frac{d^3 k}{(2\pi)^3} \big[(\Gamma^0+\Gamma)^{\lambda\mu\tau} D_{\lambda\lambda^\prime}(K) (\Gamma^0+\Gamma)^{\lambda^\prime\nu\tau^\prime} D_{\tau\tau^\prime}(R)
+(M^0 + M)^{\mu\nu\lambda\sigma} D_{\lambda\sigma}(K)\big].
\eea
The order of the momentum variables is $(K,Q,R)$ for the 3-point functions and $(Q,-Q,K,-K)$ for the 4-point functions.
Eq.~(\ref{integrand}) is to be understood as restricted to the static case
and small momentum anisotropy such that the prefactor of the integral is simply the
temperature $T$ of the isotropic distribution function (\ref{neq}).

\section{Integrand for $\alpha_{\rm NLO}$}

We calculate the NLO contribution to $\alpha$ which is obtained from (\ref{pi-invert}) as:
\bea
\alpha_{\rm NLO} = P^3_{ij}(q)\Pi_{ij}(q)\,,
\eea 
with $\Pi_{ij}(q)$ as given in (\ref{integrand}). 
The resulting integrand can be rewritten in a comparatively compact form. The method is similar to the calculations of the integrands that give the NLO equilibrium gauge boson polarization \cite{Braaten:1990it}, and the NLO equilibrium fermion self energy \cite{Braaten:1992gd,Kobes:1992ys,Carrington:2006gb}. The basic procedure is as follows. We substitute the static propagator as given in Eq. (\ref{static-prop}). Then we use the Ward identities in Eqs. (\ref{wi-full}) and (\ref{wi-tad}). This produces factors proportional to components of the HL polarization tensor in the numerator. These factors can be expressed as pieces of inverse propagators, which cancel with the corresponding components of the original propagators. This procedure allows us to identify cancellations between various pieces of the first two diagrams shown in Fig. \ref{staticSE}. We divide the result into several types of contributions: (i) terms in which all factors of the HL propagator have cancelled; (ii) tadpole-like terms which contain only one propagator; (iii) bubble-like terms that contain two propagators and no HL vertices; (iv) bubble-like terms that contain two propagators and one HL vertex; and (v) bubble-like terms that contain two propagators and two HL vertices. We thus write
\bea
\alpha_{\rm NLO} = \frac{1}{2} \,g^2\, C_A \,\delta_{ab} \,T\, \int \frac{d^3 k}{(2\pi)^3} \;[\, \sum_{i=1}^5 I^{(i)}\,]
\eea
where the result for the various contributions to the integrand $I^{(i)}$ are given below. 
To compactify the expression we use the notation:
\bea
&& {\cal P}_{kq} f(k,q):=f(k,q)+f(q,k)  \\
&& {\cal R}_{kqr} f(k,q,r):=f(k,q,r)+f(q,r,k)+f(r,k,q) \nonumber\\
&&\text{F}(k,n_k,\mu,\nu)=g_{\text{$\mu $3}} g_{\text{$\nu $3}} F_D(k)-\left(g_{\text{$\mu $3}} g_{\text{$\nu $0}}+g_{\text{$\mu $0}} g_{\text{$\nu $3}}\right)
   F_B(k)+g_{\text{$\mu $0}} g_{\text{$\nu $0}} F_C(k) \nonumber
\eea
The separate integrands then read:

\noindent (i) terms in which all factors of the HL propagator have cancelled:
\bea
\label{type-1}
I^{(1)}=-\frac{2 \left(q^2+r\cdot k \right)}{k^2 r^2}+\frac{2(2 r\cdot k-5 k_3 r_3)}{k^2 r^2 n_q^2}
\eea
   
\noindent (ii) tadpole-like terms:
 
 \bea
\label{tad-like}
I^{(2)}=&& ~ 4 G_A(k){\cal P}_{kq} \left(-\frac{\left(2 r^2- q\cdot k \right)}{r^2}-\frac{2 \left(k_3 q_3- q\cdot k \right)}{r^2 n_k^2
   n_q^2}+\frac{\left(2 k^2+q^2-k_3 \left(r_3-q_3\right)\right)}{r^2 n_q^2}\right) \\
   &&+4 G_A(k){\cal P}_{kq} \left(-\frac{q\cdot k }{q^2 r^2}+\frac{\left(q\cdot k -k_3 q_3\right)}{q^2 r^2 n_q^2}-\frac{q_3
   r_3}{q^2 r^2 n_k^2}\right) \gamma _q \nonumber\\
   && -\frac{1}{2}G_A(k){\cal P}_{kq} \left(\frac{M_{3333}}{n_k^2 n_q^2}+M_{\text{iijj}}-\frac{2 M_{\text{ii33}}}{n_q^2}\right) \nonumber\\ 
   && +\frac{4 q_3 r_3 \delta _q F_B(k)}{r^2}+\left(\frac{2}{n_q^2}-10\right) F_C(k)\nonumber\\
   &&+4  \left(\frac{2 k^2 n_k^2}{r^2 n_q^2}+\frac{\left(2 q^2+r_3^2\right)}{r^2}+\frac{4 \left(q\cdot k -k_3
   q_3\right)}{r^2 n_q^2}-\frac{q_3 r_3 \gamma _q}{q^2 r^2}\right) F_D(k)\nonumber\\
   && -\frac{1}{q^2 n_q^2}\text{F}\left(k,n_k,\mu ,\nu \right) \left(\left(M_{\text{$\mu \nu $33}}-M_{\text{$\mu \nu $ii}} n_q^2\right) q^2\right)\nonumber
   \eea


\noindent (iii) bubble-like terms with no HL vertices:
\label{type-3}
   \bea
I^{(3)}=   &&-G_A(k){\cal P}_{kq} \frac{2}{q^2}k_3 q_3 \delta _k  \left(q^2 \delta _q F_C(r)-2 \gamma _q F_B(r)\right)\\
   &&-G_A(k){\cal P}_{kq}   \frac{2}{k^2 q^2}  k_3 q_3 \gamma _k \gamma _q  F_D(r)\nonumber\\
   &&+G_A(k)  {\cal P}_{kq} \frac{4 }{n_k^2} \left(n_k^2 \left(n_q^2+1\right)-2 n_q^2\right)  \left(q^2 \delta _q F_B(r)-\gamma _q F_D(r)\right)\nonumber\\
   &&+8 q^2 n_q^2 \delta _q F_B(r) \left(F_C(k)-F_D(k)\right)\nonumber\\
   &&+4 q^2 \beta _q \left(F_B(k) F_B(r)-F_C(k) F_C(r)\right)\nonumber\\
   &&+4 n_q^2 \gamma _q \left(F_D(k) F_D(r)-F_B(k) F_B(r)\right)\nonumber\\
   &&+G_A(k) G_A(r) {\cal R}_{kqr} \left(\frac{q^2}{k^2 r^2}+\frac{n_q^2 q^2}{k^2 r^2 n_k^2 n_r^2}+\frac{2 \left(r^2 \left(n_r^2-1\right)-q^2
   n_q^2\right)}{k^2 r^2 n_k^2}-\frac{2 k_3 r_3}{k^2 r^2 n_q^2}\right) \gamma _k \gamma _r \nonumber\\
   && +G_A(k) G_A(r) {\cal R}_{kqr}  \frac{4}{n_k^2 n_r^2}  \left(\left(n_k^2-1\right) n_r^2-n_q^2 \left(n_r^2-1\right)\right) \gamma _q \nonumber\\
   && -\left(q^2 n_q^2+8 r\cdot k-8 k_3 r_3\right) F_D(k) F_D(r)+2 \left(q^2+8 r\cdot k-6 k_3 r_3\right) F_B(k)
   F_B(r) \nonumber\\
&&-\frac{\left(q^2+8 r\cdot k \; n_q^2-4 k_3 r_3\right) F_C(k) F_C(r)}{n_q^2}\nonumber\\
&&+G_A(k)F_D(r) {\cal P}_{kq}\left(2\left(1-\frac{n_r^2}{n_q^2}\right) r^2+10\left(1-\frac{1}{n_q^2}\right)(k^2 n_k^2-r^2 n_r^2)\right) \nonumber \\
&& +G_A(k) G_A(r) {\cal R}_{kqr} \left(5 k^2 n_k^2\left(\frac{1}{n_q^2}+\frac{1}{n_r^2}-\frac{1}{n_q^2n_r^2}\right)-4q^2+\frac{2q^2}{n_q^2}\right)\nonumber
\eea 

 
\noindent (iv) bubble-like terms with one HL vertex:
\bea
\label{type-4}
I^{(4)}=&& ~ G_A(k) G_A(r){\cal R}_{kqr}\frac{4}{n_k^2 n_q^2}  q_3 \left(\Gamma _{333}-n_q^2 \text{Tr$\Gamma $}_3\right) \\
&&-\frac{2}{n_q^2} \text{F}\left(k,n_k,\mu ,\lambda \right) \text{F}\left(r,n_r,\nu ,\tau \right) \left(g_{\text{$\mu $0}} g_{\text{$\nu
   $0}}-g_{\text{$\mu $3}} g_{\text{$\nu $3}} n_q^2\right) \left(k_3-r_3\right) \Gamma _{\text{$\lambda \tau $3}}\nonumber\\
   &&+ G_A(k) {\cal P}_{kq}\frac{4 }{n_q^2}\text{F}\left(r,n_r,3,\lambda \right) \left(\left(q_3-r_3\right) \Gamma _{\text{$\lambda $33}}-n_q^2 q_3
   \text{tr$\Gamma $}_{\lambda }\right)\nonumber
   \eea


\noindent (v) bubble-like terms with two HL vertices:

\bea
\label{type-5}
I^{(5)}=&& ~ G_A(k) G_A(r){\cal R}_{kqr} \left(-\frac{\Gamma _{333}^2}{3 n_k^2 n_q^2 n_r^2}+\frac{1}{3} \Gamma \cdot \Gamma-\frac{\Gamma
   _3 \cdot \Gamma _3}{n_r^2}+\frac{\Gamma _{33}\cdot \Gamma _{33}}{n_k^2 n_r^2}\right)  \\
&&   -G_A(k)\text{F}\left(r,n_r,\mu ,\nu \right) {\cal P}_{kq} \left(-\frac{2 \Gamma _{\text{$\mu $3}}\cdot\Gamma _{\text{$\nu
   $3}}}{n_q^2}+\Gamma _{\mu }\cdot \Gamma _{\nu }+\frac{\Gamma _{\text{$\mu $33}} \Gamma _{\text{$\nu $33}}}{n_k^2
   n_q^2}\right) \nonumber\\
   && +\text{F}\left(k,n_k,\mu ,\nu \right) \text{F}\left(r,n_r,\lambda ,\tau \right) \left(\Gamma _{\mu \lambda }\cdot \Gamma _{\nu
   \tau }-\frac{\Gamma _{\text{$\mu \lambda $3}} \Gamma _{\text{$\nu \tau $3}}}{n_q^2}\right)\nonumber
 \eea
 
 
 \section{Imaginary part of tadpole-like terms}
 
\begin{figure}[H]
\begin{center}
\includegraphics[width=8cm]{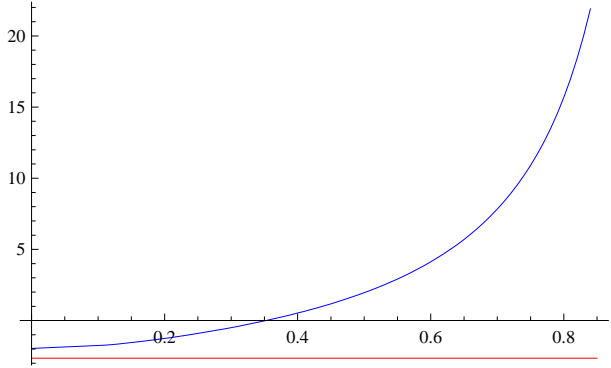}
\end{center}
\caption{(Color online) The result for $\alpha_{\rm NLO}$ as a function of $q_3/q$ for the simple tadpole (the (red) straight line) as given in Eq. (\ref{paul-tad}), and for the `full'-tadpole (the (blue) curve) as given in Eq. (\ref{tad-like}). The factor $N_c \sqrt{\xi} m_D g^2 T/(16 \pi^2)$ has been extracted.}
 \label{tadFig}
\end{figure}

We have evaluated numerically the imaginary part of the tadpole-like terms given in (\ref{tad-like}) with the HL vertices set to zero. The result is 
given in Fig. \ref{tadFig}.
We compare with the result of \cite{Romatschke:2006bb} for the tadpole diagram with a HL propagator and a bare vertex (see Appendix A):
\bea
\label{paul-tad}
{\rm Im}\;
\alpha_{\rm NLO}=-\frac{\sqrt{3}}{1024 \pi } g^2 T \sqrt{\xi } m_D N_c \left(56+5 \sqrt{2} \log \left(3-2 \sqrt{2}\right)-5
   \sqrt{2} \log \left(3+2 \sqrt{2}\right)\right) 
   \eea
\par

By considering only the tadpole diagram with a HL propagator but a bare vertex,
Ref.~\cite{Romatschke:2006bb} has obtained an imaginary part Im $\alpha$
which is of the order $g m_D^2$ and strictly negative.
Our result, which includes also contributions from the bubble diagram
with HL vertices where the latter have cancelled one of the two HL propagators,
shows that this finding is not generic as far as the sign of the result is
concerned. Evidently, the complete static NLO gauge boson self energy contains
positive as well as negative contributions to the imaginary part, and
it remains to be seen whether a complete numerical evaluation of the
expressions that we have presented here will lead to a nonzero result.

\section{Conclusions}

The main result of this paper is a relatively simple 
result for the integrand that gives the complete NLO contribution to the $\alpha$ component of the gluon self energy, in the static limit. We have calculated analytically the contribution to the imaginary part from `tadpole-like' terms, by which we mean contributions which involve only a single HL propagator after Ward identities have been used to reduce the diagram involving two HL propagators and two HL 3-vertices. The result is of the order predicted in \cite{Romatschke:2006bb}, but the tadpole-like contribution is not strictly negative.

Indeed, from previous experience with equilibrium hard thermal loop calculations, one would expect cancellations between the various contributions to the full integral. For the equilibrium case, the KMS conditions actually guarantee that the imaginary part of the self energy is zero in the static limit, at all orders. Out of equilibrium, the result must be odd in the frequency, which means that any nonzero result must contain a discontinuity at vanishing frequency. The existence of this discontinuity at
the order $gm_D^2$ has been conjectured in \cite{Romatschke:2006bb} and
used for a perturbative estimate of the order of magnitude of
jet quenching and momentum broadening parameters
\cite{Romatschke:2006bb,Baier:2008js}.
In order to determine whether there is in fact a non-zero contribution to the imaginary part of the complete static gluon self-energy at next-to-leading, the full integral, as given in Eqs. (\ref{tad-like})--(\ref{type-5}), must be evaluated numerically. This calculation, which involves some nested integrations with highly nontrivial
integrands, is in progress.

\appendix
\section{Comparison with 
Ref. \cite{Romatschke:2006bb}}\label{AppendixA}

In this Appendix we point out a few errors that occurred in
the calculation of the tadpole diagram with bare vertex which was
presented in the Appendix of Ref.~\cite{Romatschke:2006bb}. The main mistake is the incorrect 
assumption \footnote{We thank Paul Romatschke for discussions of this point.} of a relation of the form 
$\delta^{ij}\Pi^{ij} \propto n_q^i n_q^j\Pi^{ij}$
which leads to the incorrect simplification:
\bea
\label{paul-use1}
&&\beta = 0\,\\
\label{paul-use2}
&&2\alpha+\gamma = \delta_{ij}\Pi_{ij}\,.
\eea
However, the results given in Ref.~\cite{Romatschke:2006bb} do not satisfy (\ref{paul-use2}) because of typographical errors. 
We write below the results from \cite{Romatschke:2006bb} for $\alpha_{\rm NLO}$ and  $\gamma_{\rm NLO}$ from Eq. (A3) of that paper, with an extra factor $\sqrt{2}$ in front of the logs, and an extra factor (-1) in front of the result for $\gamma_{\rm NLO}$. These extra factors are written in square brackets. 
\bea\label{paul-alphagamma}
&& {\rm Im}\;\alpha_{\rm NLO\mbox{\cite{Romatschke:2006bb}}} = -\frac{1}{16 \sqrt{3} \pi }g^2 T \sqrt{\xi } m_D N_c\left(2-[\sqrt{2}]\frac{1}{4}\log \left(3+2 \sqrt{2}\right)\right) 
   \left(2-\frac{q_3^2}{q^2}\right)\\
   && {\rm Im}\;\gamma_{\rm NLO\mbox{\cite{Romatschke:2006bb}}}=[-1]\frac{1}{8
   \sqrt{3} \pi  q^2}g^2 T \sqrt{\xi }m_D N_c q_3^2 \left(2-[\sqrt{2}]\frac{1}{4}\log \left(3+2 \sqrt{2}\right)\right)  \nonumber
   \eea
These `adjusted' expressions satisfy Eq.~(\ref{paul-use2}). However Eq.~(\ref{paul-use1}) is not valid (not even at leading order).


\end{document}